\documentclass[reprint,aps,prd,superscriptaddress,preprintnumbers,nofootinbib]{revtex4-2}

\usepackage{amsfonts}
\usepackage{amsmath}
\usepackage{amsthm}
\usepackage{amssymb}
\usepackage{array}
\usepackage{dcolumn}
\usepackage[svgnames]{xcolor}
\usepackage[hidelinks]{hyperref}
\hypersetup{
    colorlinks=true,
    linkcolor=NavyBlue,
    filecolor=NavyBlue,
    urlcolor=NavyBlue,
    citecolor=NavyBlue,
    }

\usepackage{graphics}
\usepackage{tikz}
\usepackage{graphicx}
\usetikzlibrary{positioning}
\usepackage{adjustbox}
\usepackage{gensymb}
\usepackage{breqn}
\usepackage{braket}
\usepackage{enumitem}
\usepackage{gensymb}

\def\be{\begin{equation}}
\def\ee{\end{equation}}
\def\bea{\begin{eqnarray}}
\def\eea{\end{eqnarray}}

\newcommand{\gsim}{\lower.7ex\hbox{$\;\stackrel{\textstyle>}{\sim}\;$}}
\newcommand{\lsim}{\lower.7ex\hbox{$\;\stackrel{\textstyle<}{\sim}\;$}}

\begin{document}

\hfill \preprint{MI-HET-829, UH511-1329-2022, CETUP-2023-014}

\title{Multi-messenger Probes of Asteroid Mass Primordial Black Holes: \\
Superradiance Spectroscopy, Hawking Radiation, and Microlensing}

\author{James B.~Dent} 
\email{jbdent@shsu.edu}
\affiliation{Department of Physics$,$~ Sam ~Houston~ State~ University$,$~ Huntsville$,$~ TX~ 77341$,$~ USA}

\author{Bhaskar Dutta}
\email{dutta@tamu.edu}
\affiliation{Mitchell Institute for Fundamental Physics and Astronomy$,$ Department of Physics  and Astronomy$,$ Texas A\&M University$,$ College Station$,$ Texas 77843$,$  USA}

\author{Tao Xu}
\email{tao.xu@ou.edu}
\affiliation{Homer L. Dodge Department of Physics and Astronomy$,$ University of Oklahoma$,$ Norman$,$ OK 73019$,$ USA}

\begin{abstract}
Superradiance provides a unique opportunity for investigating dark sectors as well as primordial black holes, which themselves are candidates for dark matter over a wide mass range. Using axion-like particles as an example, we show that line signals emerging from a superradiated axion cloud combined with black hole Hawking radiation in extragalactic and galactic halos, along with microlensing observations lead to complementary constraints on parameter space combinations including the axion-photon coupling, axion mass, black hole mass, and its dark matter fraction, $f_{\textnormal{PBH}}$. For the asteroid mass range $\sim10^{16}-10^{22}~\textnormal{g}$, where primordial black holes can provide the totality of dark matter, we demonstrate that ongoing and upcoming observations such as SXI, JWST, and AMEGO-X will be sensitive to possible line and continuum signals, respectively, providing probes of previously inaccessible regions of $f_{\textnormal{PBH}}$ parameter space. Further complementarity from a stochastic gravitational-wave background emerging from the black hole formation mechanism is also considered. 
\end{abstract}

\maketitle

\newpage

{\bf \emph{Introduction.}} Primordial black holes (PBHs) are black holes theorized to form not from the endpoint of stellar evolution processes, but due to cosmological mechanisms in the early universe (for reviews of PBHs see~\cite{Carr:2020gox,Carr:2020xqk,Carr:2021bzv,Green:2020jor}). Such formation mechanisms include the collapse of large density fluctuations in the early universe (typically due to a feature in the inflationary potential)~\cite{1975ApJ...201....1C, Silk:1986vc, Ivanov:1994pa, Yokoyama:1995ex, Garcia-Bellido:1996mdl, Ivanov:1997ia, Kawasaki:1997ju, Garcia-Bellido:2017mdw}, particle trapping following a first order phase transition~\cite{Kawana:2021tde,Liu:2021svg,Baker:2021nyl,Baker:2021sno}, collapse from topological defects such as strings and domain walls, and scalar condensation~\cite{Cotner:2016cvr,Cotner:2017tir,Cotner:2018vug,Cotner:2019ykd}. 

PBHs have sparked great interest as possible dark matter (DM) candidates, and there are a tremendous variety of constraints on the fraction, $f_{\rm PBH}$, of DM composed of PBHs over a wide mass range (see the aforementioned reviews~\cite{Carr:2020gox,Carr:2020xqk,Carr:2021bzv,Green:2020jor}). Intriguingly, there remains a window roughly in the mass range $(10^{-15}$-$10^{-11})M_{\odot}$ (or $\sim10^{18-22}$~g) where PBHs could form the totality of DM. 

Uncharged black holes (BHs) are characterized by their mass, $M$, and angular momentum, $J$, with dimensionless spin parameter $a_* = J/(GM^2)$, with $G$ being the gravitational constant. For a Kerr black hole, it has long been known that angular momentum and energy can be extracted via the Penrose process~\cite{1969NCimR...1..252P,Penrose:1971uk}, or, analogously, via the amplification of wave scattering~\cite{1971JETPL..14..180Z,Misner:1972kx,Press:1972zz}. Additionally, a superradiant instability can lead to exponential growth in a quasi-bound state population of a massive bosonic field in a Kerr geometry background (for an in-depth review of superradiance see~\cite{Brito:2015oca}). This configuration resembles a gravitational atom in terms of its energy eigenvalues, which are set by the boson mass, $\mu$, and the dimensionless coupling $\alpha = \mu GM$~\cite{Damour:1976kh,Detweiler:1980uk,Zouros:1979iw,Dolan:2007mj,Baumann:2018vus}. Superradiance obtains when the gravitational coupling $\alpha \lesssim 1/2$, which translates to the Compton wavelength of the particle being roughly the size of the horizon crossing radius of the BH, $\mu \simeq 1/(GM)$. This provides a probe of bosons with masses in the range $\mu \approx (10^{-16}~{\rm eV},100~{\rm MeV})$ for black holes in the corresponding range  $M \approx (10^{6}M_{\odot},~10^{-18}M_{\odot})$

As this superradiant cloud forms it extracts energy and angular momentum from the black hole~\cite{Hawking:1971tu,East:2017ovw,Herdeiro:2021znw}, causing it to spin down until the superradiant instability, given by the condition $m\,\Omega - {\rm Re}(\omega) > 0$, where $\Omega$ is the angular frequency at the outer event horzon, $m$ is the azimuthal number, and $\omega$ is the boson's angular frequency, is saturated. Although PBHs generated from the collapse of density fluctuations in a radiation dominated era are expected to have low spin~\cite{Chiba:2017rvs,DeLuca:2019buf,Harada:2020pzb}, formations in other channels~\cite{Harada:2017fjm,Cotner:2019ykd,Eroshenko:2021sez} can produce larger spin values. In this work the spin origination mechanism is not considered, and $a_*$ will be taken as an external parameter (only minimal spin values of $a_{\ast}\gtrsim\mathcal{O}(10^{-3})$ are required for superradiance to occur in the $M$ ranges considered in this work~\cite{Unal:2023yxt}, though larger spin can produce more pronounced phenomenology).

Superradiance acts as a unique probe of Beyond Standard Model (BSM) physics due to its existence being predicated on the gravitational coupling rather than any hypothesized interaction between BSM and SM fields. This leads to the attractive possibility of probing particles which couple so weakly to the SM that they could not conceivably be produced in terrestrial experiments, or their cosmic abundance may be so vanishingly small that their detection via standard direct or indirect mechanisms is infeasible. 
The bosonic cloud could provide interesting observables such as a gap in the Regge plane (BH spin vs.~BH mass plane) and quasi-monochromatic gravitational waves (GWs), whose frequency is set by the boson mass ~\cite{Arvanitaki:2009fg,Arvanitaki:2014wva,Brito:2017wnc,Brito:2017zvb,Rosa:2017ury,Ghosh:2018gaw,Siemonsen:2019ebd,Ferraz:2020zgi,Tsukada:2020lgt,Yuan:2021ebu,LIGOScientific:2021rnv}.

If the superradiated particles \emph{do} have SM interactions of a size which can be probed experimentally, then new avenues of complementarity with existing approaches may be opened. For example, one could probe bosonic elements of dark sectors that couple to the SM as these particles could be superradiated and then produce observable signatures through their SM interactions. In this work we take axions or axion-like particles (ALPs), a well-motivated and well-studied BSM sector~\cite{Peccei:1977hh,Weinberg:1977ma,Wilczek:1977pj}, as our dark sector particle and examine possible signals from superradiated ALP clouds. To which SM particles and with what strength ALPs couple is model dependent (for reviews on ALP models and searches, see for example~\cite{DiLuzio:2020wdo}). However, an ALP-photon coupling, $g_{a\gamma\gamma}$, has been a mainstay of ALP searches, as it arises naturally in attempts at solving the strong CP problem~\cite{Kim:1979if,Shifman:1979if,Zhitnitsky:1980tq,Dine:1981rt}, and it is this coupling that will be explored in the present work via line signals from ALP decays to photons. 

BHs have a thermal character and radiate with a temperature inversely proportional to their mass, leading to a finite BH lifetime proportional to $M^3$~\cite{Hawking:1975vcx}. This places a bound of roughly $M \gtrsim 10^{15}$~g on the mass of PBHs that could still exist today. The products of such Hawking radiation can also be searched for, leading to current bounds on and future opportunities relating to possible PBH population parameters such as their mass and spin distributions along with their overall abundance~\cite{Clark:2016nst,Poulin:2016anj,Clark:2018ghm, Arbey:2019vqx,Coogan:2020tuf,Ray:2021mxu,Carr:2009jm}.

The redshift of the onset of superradiance affects the constraints on the PBH population, and is determined by the gravitational coupling $\alpha$. The number of ALPs in the cloud, $N_a$, is affected by $M_{\rm PBH}$, their initial spin, $a_{\ast,i}$, and their angular momentum loss, as well as their self-interaction strength, governed by $g_{a\gamma\gamma} = \alpha_{\rm EM}/(2\pi f_a)$, where $\alpha_{\rm EM}$ is the electromagnetic fine structure constant and $f_{a}$ is the ALP decay constant. The effects of self-interaction are the subject of ongoing research~\cite{Baryakhtar:2020gao,Branco:2023frw}, and in our case they manifest in two ways. A stronger self-interaction can i) quench the superradiant growth causing a decreased population of ALPs and ii) create faster decay time scales to photons.

An interesting interplay of BH processes is that Hawking radiation from spinning BHs tends to produce a harder spectrum of larger amplitude than their non-spinning counterparts, but the growth of a superradiant cloud will spin-down the BH as it extracts angular momentum, thereby reducing this effect. This leads to complementarity between Hawking radiation and superradiance for $M_{\rm PBH}\lesssim 10^{19}~{\rm g}$ where Hawking radiation will be produced in the $\sim$~MeV range while photon line signals of $\sim$~keV energy arise from decay of the superradiated ALPs. For $M_{\rm PBH} \gtrsim 10^{21}~{\rm g}$ a direct Hawking radiation observation is no longer operable, but microlensing (ML) constraints can provide additional complementarity between ML and line signals from the corresponding $\sim$~eV ALPs (there are possible signatures such as fast radio bursts, kilonovae, or gamma-ray bursts from PBH interactions with other compact objects such as white dwarfs and neutron stars whose future observation  could also probe the open asteroid-mass region~\cite{Fuller:2017uyd,Takhistov:2017bpt,Takhistov:2017nmt}).

In this work we introduce a novel and wide-ranging set of complementary probes from the combination of  galactic Hawking radiation, extragalactic Hawking radiation, and a monoenergetic photon line signal from ALP decay in the superradiant cloud. This allows us to project probes of open parameter space in the $f_{\rm PBH}-M_{\rm PBH}$ plane for a monochromatic PBH mass distribution, finding complementarity between upcoming MeV-sky searches for Hawking radiation and keV-sky X-ray line searches for $m_a\simeq~{\rm keV}$ and $M_{\rm PBH} \simeq 10^{16}-few\times10^{18}~{\rm g}$, and between ML constraints and current and future JWST line searches in the eV-sky for $m_a\simeq~{\rm eV}$ and $M_{\rm PBH} \simeq 10^{21}-10^{22}~{\rm g}$. This complementarity is also mapped into new regions of the $g_{a\gamma\gamma}-m_a$ ALP parameter space. Additionally, we comment on further complementarity through the possibility of detecting a stochastic GW background associated with the formation of the PBH.


{\bf{\emph{Superradiant ALP Production.}}} In the absence of self-interactions, the number of ALPs in the cloud is $N_a = G M^2_{\rm{PBH}} \Delta a_{\ast}/m$. When there is no quenching from ALP self-interactions, we assume $\Delta a_{\ast} = a_{\ast,i}$ such that the PBH loses all its angular momentum to that of the non-relativistic ALP cloud. As we adopt a monochromatic mass distribution with $M_{\rm PBH}$ values that are too small for accretion to have an appreciable impact~\cite{DeLuca:2020bjf}, the evolution of PBH mass and spin are only determined by the Hawking radiation and superradiance processes. To include self-interaction effects, we will consider a two-state ALP cloud system similar to that in~\cite{Branco:2023frw}, which is populated first in the $(n,\ell,m) = (2,1,1)$ state, followed by the $(3,2,2)$ state due to their relative growth rates. Including interactions between the ALPs will alter the population of each state as interactions in the $n = 2$ state can cause an ALP to move to the $n=3$ state while another drops into the PBH $ |211\rangle + |211\rangle \rightarrow |0\rangle + |322\rangle$, while interactions in the $n=3$ state can cause an ALP to drop down to the $n=2$ state while another is ionized $|322\rangle + |322\rangle \rightarrow |211\rangle + |\infty\rangle$ with $|\infty\rangle$ representing the ionized state. For a self-interaction of the form
\bea
\mathcal{L}_{int} \supset
\frac{m_a^2}{f_a^2}\frac{a^4}{4!} \equiv \frac{\lambda}{4!}a^4,
\eea
its effects are controlled by the parameter $\lambda$ in determining the level transition rates and the equilibrium occupation number of the $n=2$ and $n=3$ states. Due to the suppression of $N_a$ from the self-interaction, the ALP cloud growth saturates once the equilibrium number density is reached, leading to a lower PBH spin extraction rate with larger $\lambda$ values. 

The proper decay width of the ALP to two photons is 
\bea
\Gamma_a=\frac{g^2_{a\gamma\gamma}~m^3_a}{64\pi}.
\eea
We focus on di-photon decay as the direct channel to detect the superradiance process. To produce enough ALPs in the cloud, we require that the exponential growth of the $|211\rangle$ state occurs at a time scale smaller than the age of the Universe.
For a given $M_{\textnormal{PBH}}$, this provides a window in the range of ALP masses which satisfies the superradiance requirement for different $a_{\ast,i}$ values. 
We checked that, for the benchmark points of this study, the superradiated ALP only contributes to small fractions of the DM relic density, thus cosmological constraints on their decay are weak due to their negligible abundance.

\begin{figure}
        \centering
        \includegraphics[width=\linewidth]{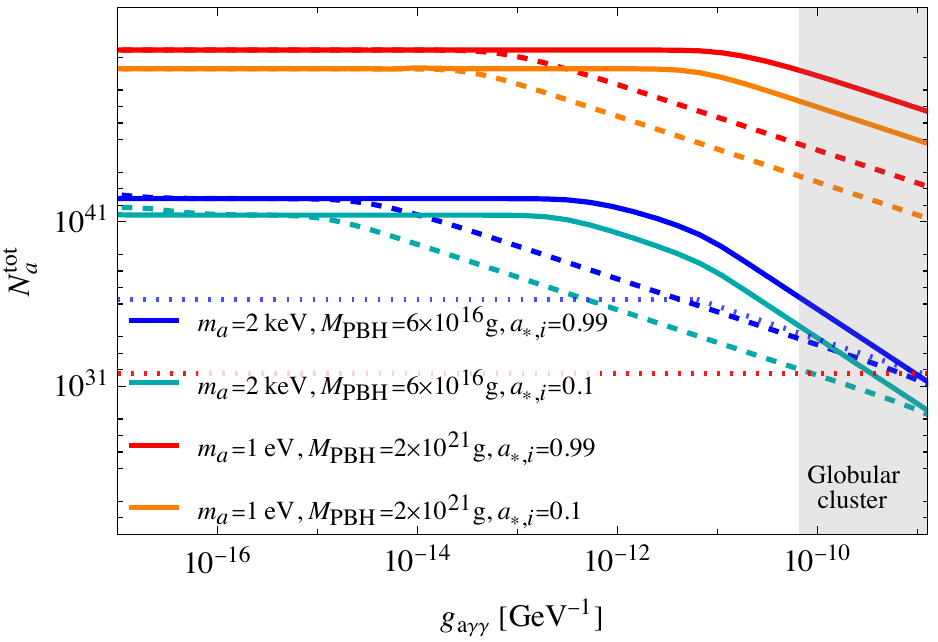}
        \caption{The total numbers of ALPs $N_{a}^{\textnormal{tot}}$  that a PBH can produce as a function of ALP self-interaction strength represented with the $g_{a\gamma\gamma}$ coupling. The $N_{a}^{\textnormal{tot}}$ in the present universe (solid), and at the CMB epoch $z=1100$ (dashed)  show the effect of the replenishment process that ionize the ALP cloud. The shaded regions show the excluded $g_{a\gamma\gamma}$ regions for ALP mass from globular cluster (gray)~\cite{Ayala:2014pea,Dolan:2022kul}. The axion production from Hawking radiation is shown in dotted lines for $M_{\rm{PBH}} = 6\times10^{16}$~g (blue) and $M_{\rm{PBH}} = 2\times10^{21}$~g (red) with $a_{\star,i}=0.99$.} 
    \label{fig:Ninfga} 
\end{figure}

In Fig.~\ref{fig:Ninfga} we show the net number of  ALPs, $N_{a}^{\textnormal{tot}}$ (including the bound and ionized states), as a function of $g_{a\gamma\gamma}$ for two ALP masses: $m_a= 1~{\rm eV}$ and $2~{\rm keV}$, with corresponding $M_{\rm PBH}$ benchmark values that allow ALP superradiance.
We find that $N_{a}^{\textnormal{tot}}$ is dominated by the ionized ALPs and, as expected, for larger $M_{\rm PBH}$, we have more ALPs. The number of ALPs also depends on  $a_{*,i}$ and $f_a$ (through its relation to $g_{a\gamma\gamma}$), with larger $a_{*,i}$ and  $f_a$ (giving smaller $g_{a\gamma\gamma}$) values both contributing to an increase in $N_a$ due to larger angular momentum loss and smaller self-interactions, respectively. Although $N_a$ for individual PBHs is independent of $f_{\rm PBH}$, the combined signal from superradiant decays will increase with increasing $f_{\rm PBH}$.  The horizontal parts of the lines correspond to small self-interactions, while the parts beyond the knee regions of the lines correspond to larger self-interactions and a loss of superradiance. During earlier times, the knee region is shifted to the left (shown by dashed lines for ALP number during the CMB era) since there was not enough time to populate the ionized levels, and a smaller self-interaction began to impact the number of ionized ALPs. Hawking radiation can also produce ALPs in the late universe~\cite{Agashe:2022phd, Jho:2022wxd}, but the abundance from evaporation shown by the dotted lines is negligible compared to superradiance of rotating PBHs for stable ALPs at cosmic time scales, as well as for unstable ALPs during the ALP lifetime. The ALP-photon interaction is mainly constrained by the Primakoff process in stellar evolution ~\cite{Ayala:2014pea,Dolan:2022kul}. The $g_{a\gamma\gamma}$ bounds obtained for keV-scale ALPs with the assumption of ALPs being all of DM are weakened for ALPs produced by PBH superradiance because their relic abundance is suppressed compared to that of DM. High-redshift constraints also need to be considered with the inclusion of a finite lifetime in large $g_{a\gamma\gamma}$ regions. We used these example scenarios from Fig.~\ref{fig:Ninfga} to calculate Hawking radiation and line signals.


{\bf{\emph{Complementarity of Superradiance with other observations.}}} The decays of ALPs at rest emerging from the superradiance cloud  in the bound and ionized states produce line signals. The line signal energy thus depends on the ALP mass scale, which in turn depends on $M_{\textnormal{PBH}}$. For example, ALPs  with $m_a\sim$ keV can be accommodated in the superradiance cloud of a PBH with $M_{\textnormal{PBH}}\sim 10^{16}$~g, while $m_a\sim$ eV ALPs require $M_{\textnormal{PBH}}\sim 10^{21}$~g. PBH masses of $\mathcal{O}(10^{16})$~g could also be detected from Hawking radiation at the upcoming AMEGO-X instrument~\cite{Caputo:2022xpx} (other proposed instruments covering a similar energy regime include e-ASTROGRAM~\cite{e-ASTROGAM:2017pxr,Coogan:2020tuf}, GECCO~\cite{Bottacini:2023hpy,Moiseev:2023zkv}, and COSI~\cite{Caputo:2022dkz}) since these PBHs could presently be producing MeV scale radiation, while $M_{\textnormal{PBH}}\sim 10^{21}$~g can be probed in microlensing observations.  $f_{\textnormal{PBH}}\lesssim 10^{-3}$ is required for $M_{\textnormal{PBH}}\sim 10^{16-17}$~g due to constraints from COMPTEL~\cite{Ray:2021mxu}, and this range is also accessible at AMEGO-X. Neither $f_{\textnormal{PBH}}$ nor $M_{\textnormal{PBH}}$ is currently constrained in the $10^{17}-10^{21}$~g range. However, line signals from superradiance could provide an opportunity to probe some parts of these unconstrained regions as well (some other possibilities can be found in~\cite{Fuller:2017uyd,Takhistov:2017bpt,Takhistov:2017nmt}). 

The gamma-ray flux from Hawking radiation for AMEGO-X
is calculated using
\bea
\frac{d\Phi}{dE_\gamma}= J_{D}\frac{\Delta \Omega}{4 \pi} \, \frac{f_{\rm PBH} }{M_{\rm PBH}} \, \frac{d^2 N_{\gamma}}{dE_{\gamma} dt}.
\eea
The ${d^2 N_{\gamma}}/{dE_{\gamma} dt}$ expression is the number of photons produced by Hawking radiation per unit energy and time from a single PBH. The region-of-interest is chosen to be $R<5^{\circ}$ for the galactic center observation with $\Delta \Omega=2.39 \times 10^{-2}~{\rm sr}$, and the corresponding J-factor for PBH distribution is $J_D=1.597 \times 10^{26}~{\rm MeV} {\rm cm}^{-2} {\rm sr}^{-1}$.

The flux of the line signal from the decay of ALPs produced by superradiance is calculated with
\bea
\frac{d\Phi_{\rm sr}}{d E_{\gamma}}=J_D \, \frac{\Delta\Omega}{4\pi}\, \frac{f_{\rm PBH} \, N^{\rm tot}_a }{M_{\rm PBH}} \, \Gamma_a \, \left(\frac{df}{dE_{\gamma}}\star W\right).
\eea
Here $df/dE_{\gamma}$ is the ALP decay spectrum to be convolved with the Gaussian detector response $W$ accounting for the energy resolution to the decay line. We use blank-sky observations to search for the ALP decay signal from the Milky Way halo~\cite{Dessert:2018qih, Foster:2021ngm, Dessert:2023vyl}. For $m_a=1~{\rm eV}$, we use public JWST Near-Infrared Spectrograph data with $J_D\Delta\Omega=2.315\times 10^{25}~{\rm MeV} {\rm cm}^{-2}$. For $m_a=2~{\rm keV}$, we use projected SXI sensitivity with $J_D\Delta\Omega=1.0 \times 10^{25}~{\rm MeV} {\rm cm}^{-2}$~\cite{Thorpe-Morgan:2020rwc, obrien2018soft, THESEUS:2017qvx}. Blank-sky observations with JWST~\cite{Janish:2023kvi,Roy:2023omw} and SXI~\cite{Thorpe-Morgan:2020rwc}, and spectroscopic searches of dwarf spheroidal galaxies with WINERED~\cite{Bessho:2022yyu, Yin:2024lla} have also been used to probe the decay of light DM.

\begin{figure}
    \centering
        \includegraphics[width=\linewidth]{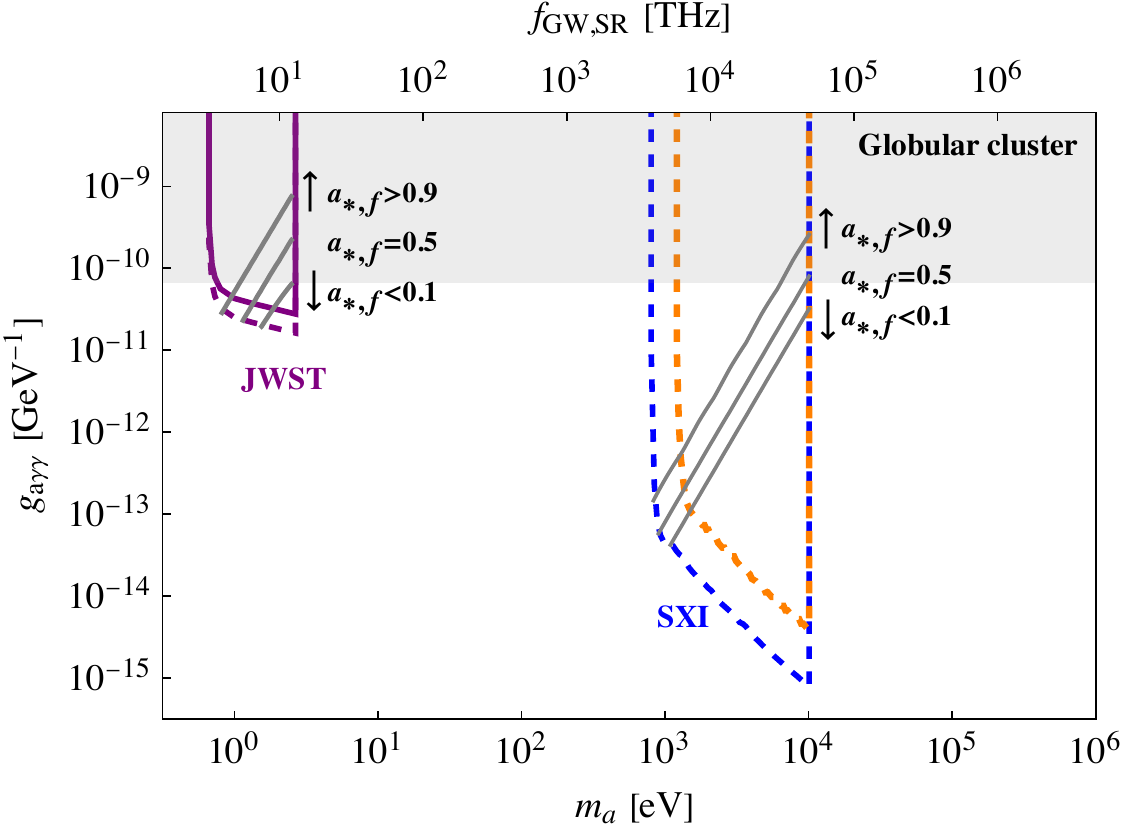}
        \caption{Projected sensitivity to the ALP parameter space 
        from superradiance cloud decay. See text for the description.
        }
    \label{fig:gama}
\end{figure}

It is interesting to examine the $g_{a\gamma\gamma}$-$m_a$ ALP parameter space using the line signals at the ongoing JWST and the upcoming SXI. In Fig.~\ref{fig:gama}, we show their $2\sigma$ reach for $m_a\sim~\textnormal{eV}-\mathcal{O}(10)$~keV. We note that JWST and SXI probe $g_{a\gamma\gamma}$ values beyond the current globular cluster constraints. For the SXI projection, the orange and blue dashed lines correspond to $f_{\textnormal{PBH}}=8\times 10^{-5}$ and $2\times10^{-3}$, respectively, for $M_{\textnormal{\textnormal{PBH}}}=6\times 10^{16}$~g. Higher $f_{\textnormal{PBH}}$ values allow for probes of smaller $g_{a\gamma\gamma}$ values due to greater total ALP and subsequent gamma-ray production from the greater PBH population. However, the Hawking radiation also constrains larger values of $f_{\textnormal{PBH}}$. The range of final PBH spin, $a_{\ast,f}$, is shown from 0.9 to 0.1 in the parameter region within future SXI sensitivity. The current JWST constraint is shown by the solid purple line, while the dashed purple line describes a future projection after 15 years of running~\cite{Janish:2023kvi}. The $a_{\ast,f}$ value is also shown in the range $0.1-0.9$ in this case. Both lines are drawn for $f_{\textnormal{PBH}}=1$ and $M_{\textnormal{PBH}}=2\times10^{21}$~g. 

On the horizontal axis on the top of Fig.~\ref{fig:gama} we also display the frequency of the GWs  which can be generated from the ALPs in the superradiance cloud annihilating into gravitons. Such a high frequency spectrum related to this PBH mass range would be challenging to access at currently proposed GW detectors~\cite{Aggarwal:2020olq,Domcke:2023qle,Kahn:2023mrj}.

\begin{figure}
    \centering
        \includegraphics[width=0.95\linewidth]{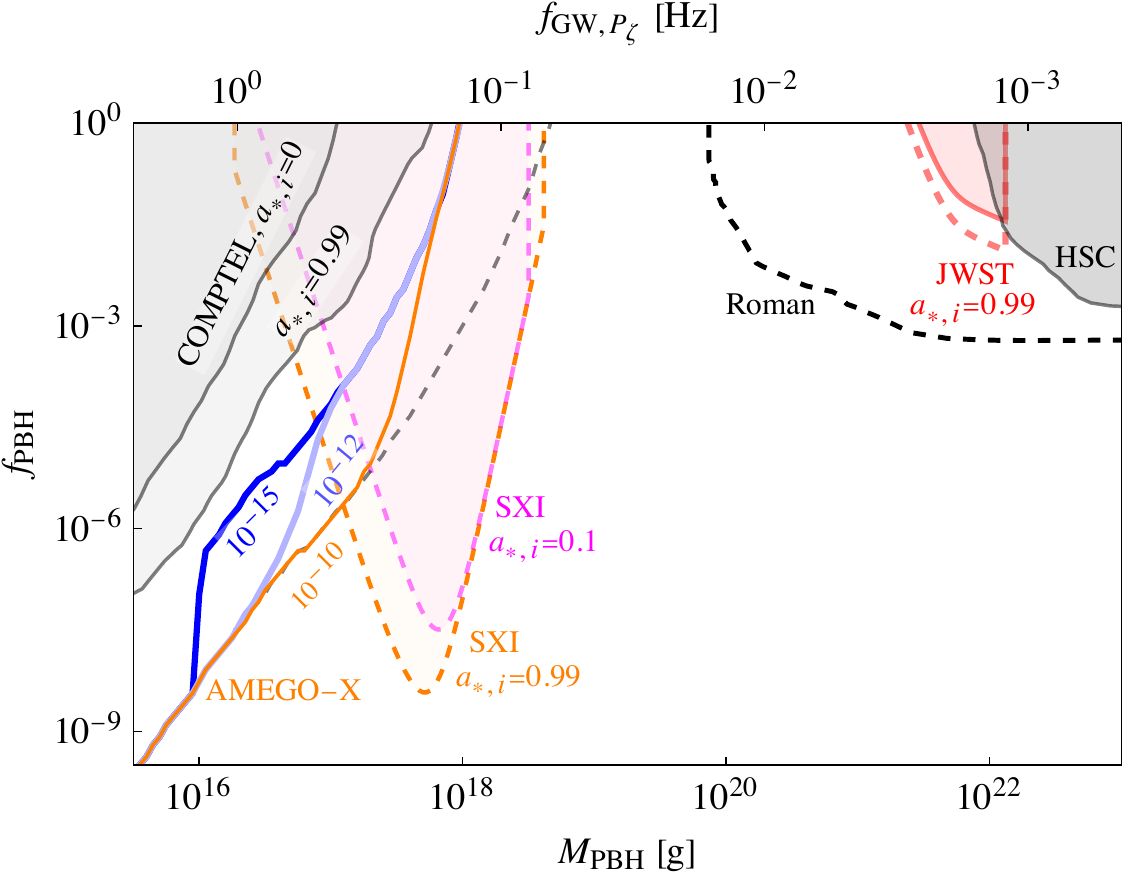}
        \caption{$f_{\rm PBH}$ vs $M_{\rm PBH}$ parameter space after including the existing and upcoming constraints. See text for the description.}
    \label{fig:fPBHMPBH1}
\end{figure}

In Fig.~\ref{fig:fPBHMPBH1}, we show the $2\sigma$ constraints in the $f_{\rm PBH}$ vs. $M_{\rm PBH}$ parameter space for our benchmark values of $g_{a\gamma\gamma}$ and $m_a$.  On the left side of the plot, constraints emerge from Hawking evaporation projected for AMEGO-X observations of galactic center gamma-rays, and line signals projected for SXI due to superradiated ALPs decaying to X-rays. We show the existing COMPTEL constraint on the PBH parameters, derived from Hawking radiation in galactic center observations, for both spinning ($a_{\ast,i}=0.99$) and non-spinning ($a_{\ast,i}=0$) PBHs in the gray shaded regions on the left~\cite{Coogan:2020tuf, Kappadath:1998PhDT}. The dashed gray line provides a projection for the reach of AMEGO-X for $a_{\ast,i} = 0.99$ with no superradiance present. In contrast, the color curves show that the Hawking radiation sensitivity is weakened when superradiance reduces the radiation power, as seen by the solid orange ($g_{a\gamma\gamma}=10^{-10}$ GeV$^{-1}$), light blue ($g_{a\gamma\gamma}=10^{-12}$ GeV$^{-1}$), and solid blue ($g_{a\gamma\gamma}=10^{-15}$ GeV$^{-1}$) lines when the superradiance condition is satisfied by the given PBH mass and ALP mass of $m_a=2~{\rm keV}$ for an initial spin of $a_{\ast,i}=0.99$. A complementary X-ray signal from the ALP cloud is induced in X-ray line searches shown in the shaded orange ($a_{\ast,i}=0.99$) and magenta region ($a_{\ast,i}=0.1$) for $g_{a\gamma\gamma}= 10^{-10}$ GeV$^{-1}$. We find that the line signals from superradiance can explore available PBH parameter space for a large range of $f_{\textnormal{PBH}}$ and $M_{\textnormal{PBH}}$ values beyond the Hawking radiation constraint.

The red-shaded region on the right shows the parameter space for line searches of superradiated $m_a = 1~\textnormal{eV}$ ALP decays constrained by the current JWST data, while the red dotted line shows the future constraints from JWST.
It is interesting to note that the current reach of JWST for eV mass ALPs puts bounds on the PBH masses where $f_{\textnormal{PBH}}$ currently is unconstrained. Future microlensing results (black dotted line) will probe this parameter space. The gray-shaded region with a solid black line contour shows the current constraint from microlensing~\cite{Niikura:2017zjd, Montero-Camacho:2019jte, Smyth:2019whb, Croon:2020ouk, Bird:2022wvk}. 

The formation of PBHs~\cite{Carr:1974nx, Carr:1975qj,Hawking:1987bn,Baker:2021sno,Kawana:2021tde} can also produce astochastic GW background~\cite{Tomita:1967wkp,Ananda:2006af,Baumann:2007zm,Saito:2008jc,Kohri:2018awv,Byrnes:2018txb,Kozaczuk:2021wcl,Liu:2021svg}.  One popular formation mechanism is due to the scalar curvature perturbation, $P_\zeta$, arising from inflation which, at second-order, induces a  possibly observable stochastic GW background. Recently, a correlation between Hawking radiation and such a scalar-induced GW (SIGW) background was analyzed~\cite{Agashe:2022jgk}. Across the top of Fig.~\ref{fig:fPBHMPBH1}, we show the peak frequency of this SIGW signal across our PBH mass range (the intensity, $\Omega_{\textnormal{GW}}$, would depend on the amplitude of $P_\zeta$). The values of these frequencies show that  BBO~\cite{Corbin:2005ny} and DECIGO~\cite{Kawamura:2020pcg} can investigate $M_{\textnormal{PBH}}$ values which will be probed at AMEGO-X and SXI, while LISA~\cite{amaroseoane2017laser} will explore the PBH masses which are being probed at JWST using the superradiance line signals. For reference, the LIGO-VIRGO-KAGRA sensitivity to such SIGW signals are at GW frequencies corresponding to $M_{\textnormal{PBH}}\lesssim 10^{16}~\textnormal{g}$~\cite{KAGRA:2021vkt}.


{\bf{\emph{Conclusions.}}} Superradiance coupled with signatures such as Hawking radiation and mircrolensing provides an intriguing tool with which to study dark sector particles in the presence of spinning primordial black holes. Using ALPs and PBHs as our dark sector laboratory, we have demonstrated that  superradiance induced photon line signals, gamma-rays from Hawking radiation, and microlensing observations can probe new regions of  $M_{\textnormal{PBH}}$, $f_{\textnormal{PBH}}$, ALP mass, and ALP-photon coupling parameter spaces in a complementary way across a large span of (eV-MeV) photon energies. We showed that keV-scale line signals at the upcoming SXI and Hawking radiation spectra at AMEGO-X will probe $M_{\textnormal{PBH}}$ up to $10^{18}$~g where PBHs could provide the totality of the DM. Line signals from superradiance at SXI can probe a large region of PBH and ALP parameter space, going beyond the reach of Hawking radiation searches. Similarly, the ongoing JWST telescope exploration of eV-scale lines gives sensitivity in the  $M_{\textnormal{PBH}}\sim 10^{21}-10^{22}$~g range, which is below the mass range of current microlensing bounds, but can be seen in next generation telescopes. We also find that, in our benchmark scenarios where PBHs compose a fraction of the DM abundance that satisfies current constraints from Hawking radiation searches, $g_{a\gamma\gamma}$ can be probed to more than four orders of magnitude below the current bounds. Additionally, searches for a stochastic GW background provide another complementary probe of the $f_{\textnormal{PBH}}-M_{\textnormal{PBH}}$ space in question, assuming that the PBHs are formed from scalar curvature perturbations, $P_\zeta$, which induce a GW background at second order. Similar correlations can also be obtained for other types of PBH formation mechanisms. We find that BBO and DECIGO would be sensitive to $M_{\textnormal{PBH}}$ values which will be probed at keV-MeV sensitive observatories such as AMEGO-X and SXI, while LISA will probe the PBH masses which are being investigated at JWST using the superradiance line signals.
Overall, the correlation of these various types of signals could tell us not only about the PBH fraction of dark matter 
 and the possible existence of ALPs, but also give clues to the formation history of PBHs. 

\noindent \textbf{\textit{Acknowledgments --}} 
The authors would like to thank Elena Pinetti, Kuver Sinha, and Volodymyr Takhistov for helpful comments on a draft version of the manuscript. The authors would like to thank Nils Siemonsen and William E. East for helpful clarifications regarding the \texttt{SuperRad} software package~\cite{Siemonsen:2022yyf}, which we utilized for superradiance calculations, Ryan Janish and Elena Pinetti for details regarding JWST line searches~\cite{Janish:2023kvi}, and Regina Caputo for discussion relating to AMEGO-X. The work of BD is supported in part by the U.S.~Department of Energy Grant DE-SC0010813. JBD acknowledges support from the National Science Foundation under grant no. PHY-2112799. The work of TX is supported by the U.S.~Department of Energy Grant DE-SC0009956.

\appendix

\section{Some Elements of Hawking Radiation and Superradiance}

\noindent The dimensionless spin parameter of a rotating PBH can be defined with its angular momentum $J$ and mass $M_{\rm PBH}$, $a_{\ast}=J/(G M^2_{\rm PBH})$, or with its angular frequency $\Omega$ at the outer horizon, $a_{\ast}=2 \, \Omega \, r_{+}$. where the outer horizon is at $r_{+}=G M_{\rm PBH}(1+\sqrt{1-a^2_{\ast}})$. A Schwarzschild PBH is non-rotating with $a_{\ast}=0$ and the extremal rotating state is given by $a_{\ast}=1$. The superradiance process occurs when the condition $m\,\Omega > {\rm Re}(\omega) $ is satisfied with the choice of the ALP mass, the PBH mass, and PBH spin. The real part of the frequency for principle quantum number $n$ is 
\bea
{\rm Re}(\omega)\simeq m_a \left( 1 - \frac{\alpha^2}{2\, n^2}\right).
\eea
Again, $\alpha = m_a GM_{\rm PBH}$ is the gravitational coupling for the hydrogen-like bound state. The superradiance cloud growth rate is determined by the imaginary part of the frequency. We calculate the superradiance rate using numerical values in the \texttt{SuperRad} code~\cite{Siemonsen:2022yyf}. However, we will comment here on the slight adjustments made with respect to what is described in~\cite{Siemonsen:2022yyf} in order to produce the numerical results presented in the present work.

The growth rate of the superradiant scalar cloud is given by Eqs.~(13) and (14) of \cite{Siemonsen:2022yyf}
\begin{align}
    \omega_{I}M = \alpha^{4m+5}\left(\omega_R-m\Omega\right)2r_{+}G_{S}(a_{\star},\alpha)
\label{eq:omegaIM}
\end{align}
with $G_S$ given in Eq.~(B4) of \cite{Siemonsen:2022yyf}. The real and imaginary frequencies, $\omega_R$ and $\omega_I$, respectively, are determined using fitting functions. For $\omega_R$ the fitting formula has coefficients $\hat{a}_{p,q}$, while $\omega_I$ has coefficients $\hat{b}_{p,q}$ and $\hat{c}_{p,q}$ as discussed in their Appendix C. It was found that in order to reproduce the correct numerical results for the growth rate for the $m_S=2$ state (used for ALPs of the present study), one not only needs to use the $(p,q)$ values explicitly mentioned in Eq.(15) of~\cite{Siemonsen:2022yyf} for each expansion coefficient, but to also include the following terms in $\omega_{I}$:
\bea
-1.1948426572069112\times 10^{11}\alpha^{12}&&\nonumber\\
+2.609027546773062\times10^{12}\alpha^{13}.&&
\eea
which are provided in the file \texttt{rel_sca_cloud.py} of the \texttt{SuperRad} code and are already used for the extrapolated results in that work.\footnote{The authors would like to thank W.E. East for a helpful clarification on this point.} We also found that using $q\in\{0,1,2\}$ for $\hat{b}_{p,q}$ and $\hat{c}_{p,q}$ in the fitting function of $m_S=2$ best reproduces the numerical result. 

\begin{figure}
        \includegraphics[scale=0.45]{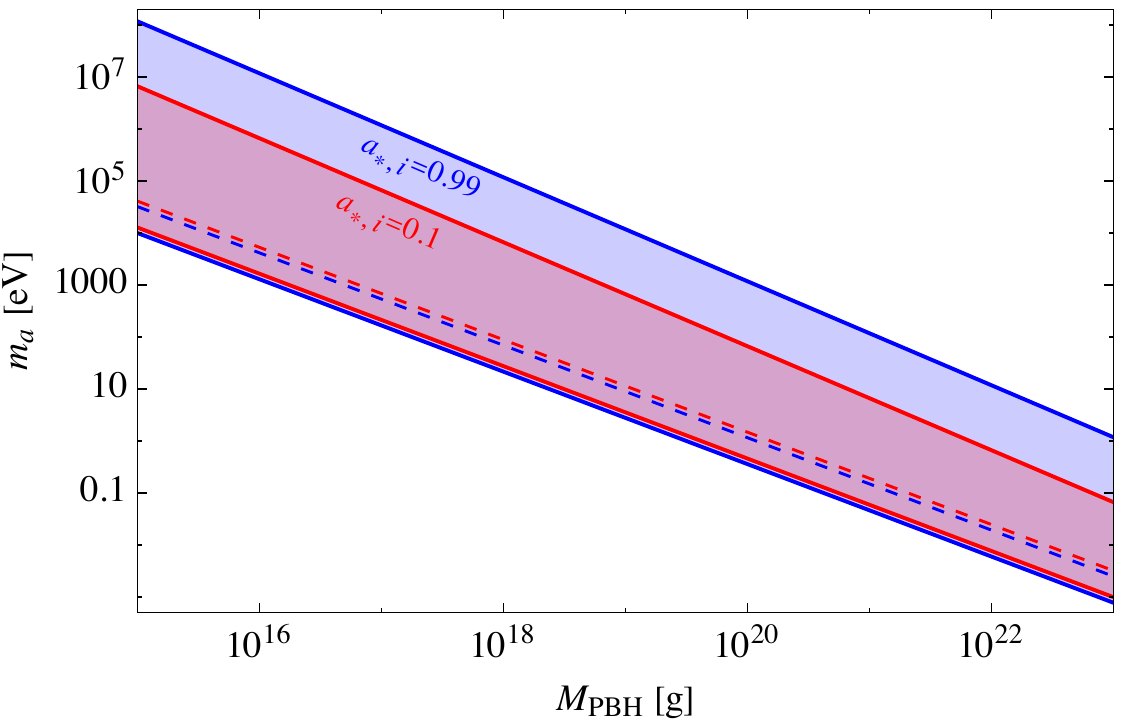}
        \caption{The shaded regions show the range of $m_a$ where the exponential growth of the $|211\rangle$ state occurs at time scale smaller than the age of the universe for initial dimensionless spin values of $a_{*,i} = 0.99$ (blue) and $a_{*,i} = 0.1$ (red). The upper edges of shaded regions are set by the superradiance condition. The lower edges are determined by the condition that the superradiance growth rate is larger than the inverse of the age of the universe. The dashed curves show the time scale of CMB, above which superradiance occurs earlier.}
    \label{fig:appendixSRtimescale} 
\end{figure}

Fig.~\ref{fig:appendixSRtimescale} shows the parameter space for the fastest growing mode, $(n,l,m)=(2,1,1)$, for initial spin $a_{\ast,i}=0.99$ (blue) and $0.1$ (red). Shaded regions show the ALP mass that can be produced in the cloud for a wide range of the PBH mass. The time scale of the superradiance process is shown for the age of the universe (solid lower edges), and CMB time (dashed curves).

\begin{figure*}
    \centering
        \includegraphics[width=0.49\linewidth]{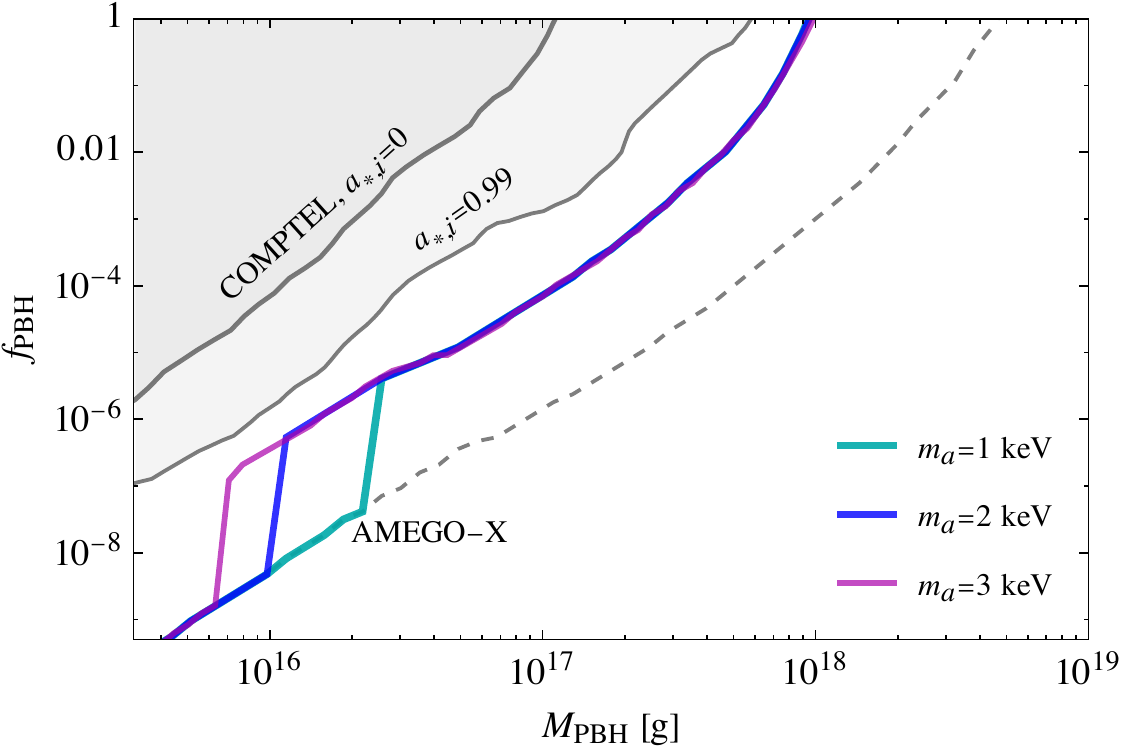}
        \includegraphics[width=0.49\linewidth]{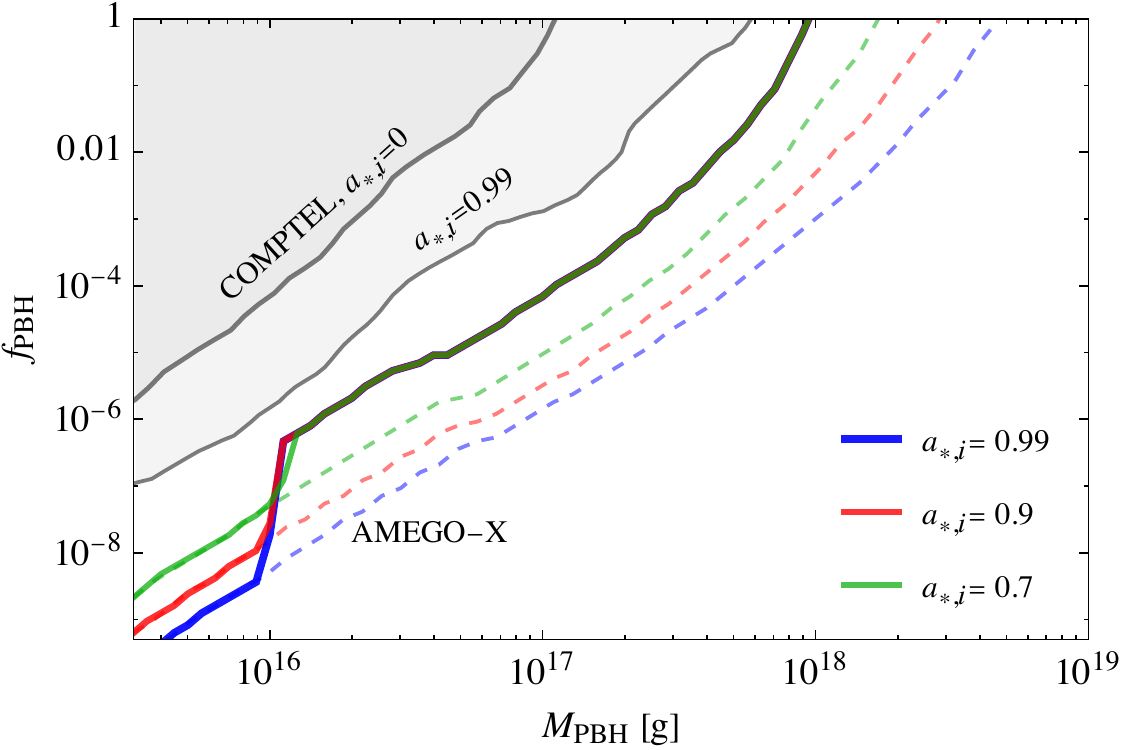}
        
        \caption{The projected $2\sigma$ sensitivity to $f_{\rm PBH}$ assuming $10$-year AMEGO-X observation of the galactic center within $5^\circ$. The gray shaded regions represent existing COMPTEL constraints on PBH Hawking radiation from galactic center observations, assuming different spin values. {\it Left}: The color curves are for the cases when the superradiance instability is triggered by an ALP of mass $m_a= 1~{\rm keV}$ (cyan), $2~{\rm keV}$ (blue), and $3~{\rm keV}$ (purple). The axion coupling is fixed to be $g_{a\gamma\gamma}=10^{-15}~{\rm GeV}^{-1}$. The PBH initial spin is chosen to be $a_{\ast,i}=0.99$. {\it Right}: The color curves are for the cases when an ALP of fixed mass $m_{a}=2~{\rm keV}$ and interaction $g_{a\gamma\gamma}= 10^{-15} {\rm GeV}^{-1}$ triggers the superradiance instability for PBHs of initial spin $a_{\ast,i}=0.99$ (blue), $0.9$ (red), and $0.7$ (green) respectively.}
    \label{fig:appendixfPBHmPBH}
\end{figure*}

Here we remark on the uncertainties in the superradiance calculation. In this work, we use correction terms in \cite{Siemonsen:2022yyf, Ferraz:2020zgi} to obtain superradiance cloud growth rate that numerically agrees with \cite{Dolan:2007mj}. In the parameter region of large $\alpha$, corresponding to heavier ALP or PBH masses, corrections to \eqref{eq:omegaIM} are needed in order to obtain the correct growth rate in the strong coupling scenario. In the parameter region of large $g_{a\gamma\gamma}$ couplings, the ALP self-interaction becomes important in determining the cloud evolution when the equilibrium axion number in each energy level is saturated. We use Boltzmann equations to calculate the superradiance axion number assuming $|211\rangle$ and $|322\rangle$ states dominate the axion population in the bound state. See \cite{Branco:2023frw} for analytical solutions of different axion self-interaction scenarios. The distribution of plasma particles around PBHs could also modify the evolution of the superradiance cloud, depending on the interaction between ALPs and the plasma. We do not include the plasma effect in our simulations, but refer readers  to~\cite{Spieksma:2023vwl} for details.

The Hawking temperature of a PBH with spin $a_{\ast}$ is 
\bea
T_{\rm PBH}=\frac{1}{4 \pi G M_{\rm PBH}}\left(\frac{\sqrt{1-a^{2}_{*}}}{1+\sqrt{1-a^{2}_{*}}}\right).
\eea
Although the Hawking temperature is lower for larger $a_{\ast}$, the Hawking radiation rate is enhanced by the PBH angular momentum as
\bea
\frac{d^2N_i}{dE dt}=\frac{g_i}{2\pi} \frac{\Gamma}{e^{E'/T_{\rm PBH}}-(-1)^{2s}}.
\eea
Here $E'=E-m\Omega$ is the effective energy of the emitted particle, and $s$ and $m$ are the spin and axial angular momenta of the emitted particle, respectively. The quantity $\Gamma$ is derived from the greybody factor of rotating PBHs. We take numerical values of $\Gamma$ as a function of the PBH spin and the particle energy normalized to the Hawking temperature from the package \texttt{BlackHawk}~\cite{Arbey:2021mbl}. To calculate the PBH Hawking radiation rate during the spin-down process, we interpolate the emission rate obtained from \texttt{BlackHawk} on the grid of $\{M_{\rm PBH}, a_{\ast}\}$. We checked the interpolated radiation rates reproduced emission rates in \texttt{BlackHawk}.

In Fig.~\ref{fig:appendixfPBHmPBH} (left) we show the spin-down effect with various superradiated particle masses (from $m_a = 1$~keV to $m_a = 3$~keV) while the coupling $g_{a\gamma\gamma}$ and initial spin $a_{*,i}$ are held fixed. We see that a larger $m_a$ value causes superradiance to occur at a smaller $M_{\textnormal{PBH}}$ value, decreasing the reach of Hawking radiation searches from a near extremal spinning PBH with accompanying harder emission spectrum (the gray dashed line), to a PBH with a drastically reduced spin and thus a softer Hawking radiation spectrum. 

In Fig.~\ref{fig:appendixfPBHmPBH} (right) we have fixed the ALP mass and coupling while allowing the initial spin to vary. For $M_{\textnormal{PBH}} \gtrsim 2\times 10^{16}$~g superradiance is operable and the PBH spins down. This produces the overlap of the solid lines in this mass range. The dashed lines show the reduction in the reach for Hawking radiation searches due to a softer spectrum for successive  spin reductions where each spin is assumed to be constant and superradiance is not present. Below $M_{\textnormal{PBH}} \simeq 2\times 10^{16}$~g superradiance is not operable and the solid curves follow the reach of Hawking radiation observations expected for PBHs of the given (constant) spins.

\section{Benchmark Example of Multi-messenger Signals}

\begin{figure*}
    \centering       
        \includegraphics[width=0.49\linewidth]{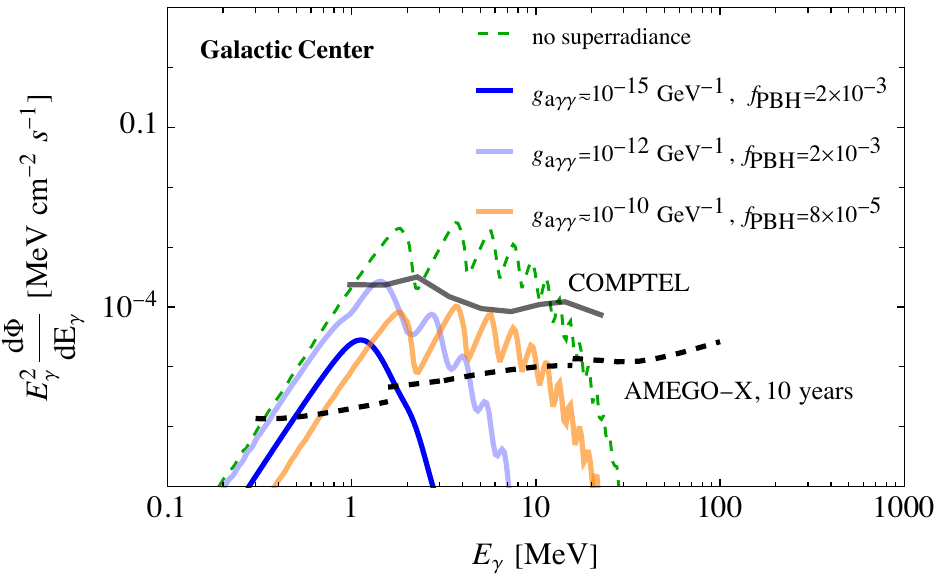}
        \includegraphics[width=0.49\linewidth]{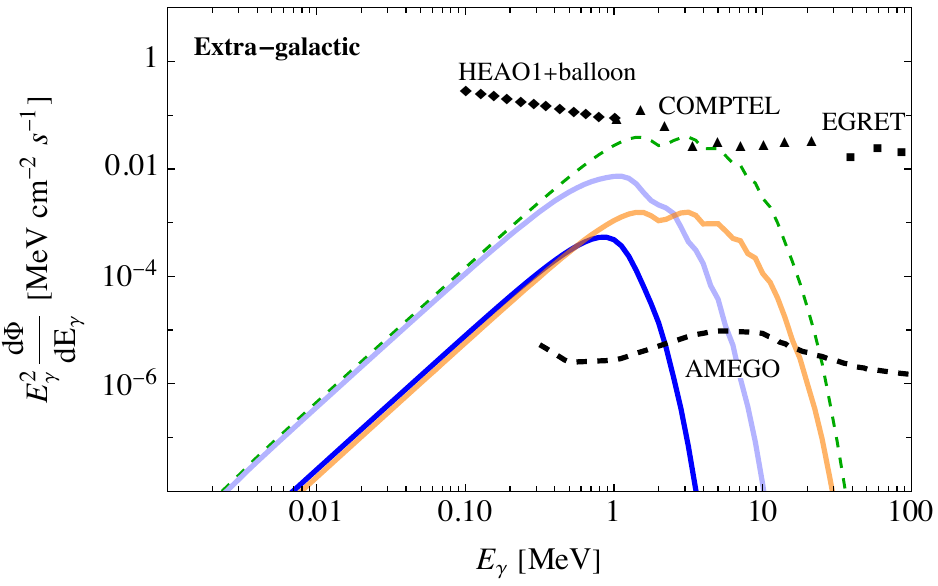}\\
        \includegraphics[width=0.49\linewidth]{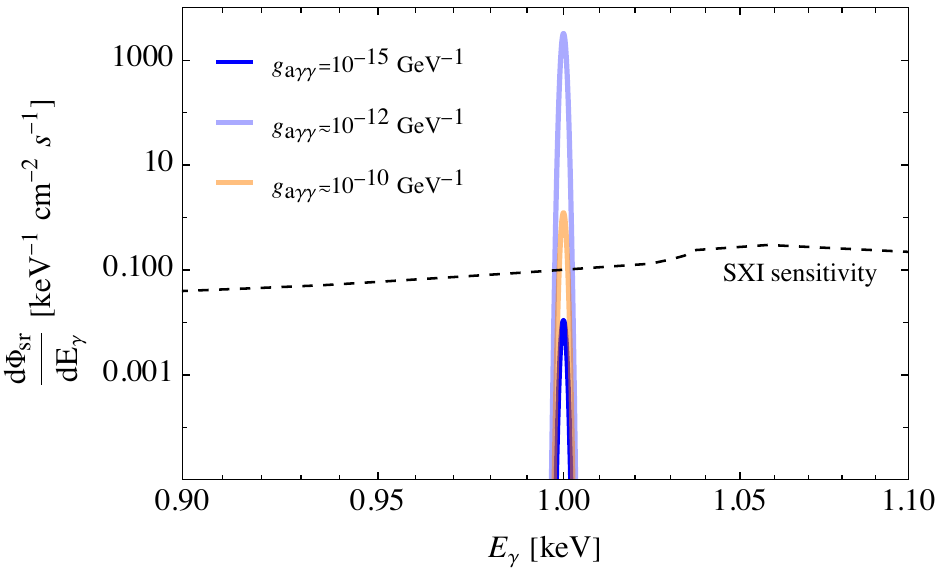}\qquad
        \includegraphics[width=0.47\linewidth]{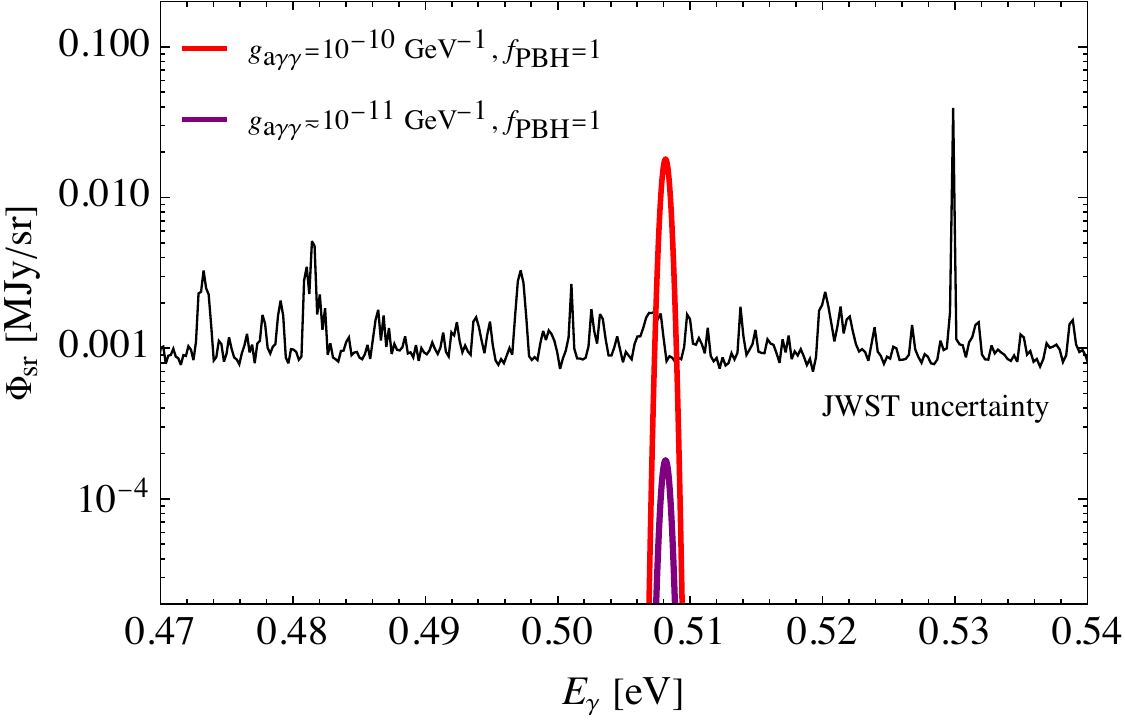}
        \caption{ (Top-left) Galactic center gamma-ray spectrum from Hawking radiation of PBH mass $M_{\rm PBH}=6\times10^{16}~{\rm g}$ and ALP mass $m_a=2~{\rm keV}$, with $g_{a\gamma\gamma}=10^{-15}~{\rm GeV}^{-1}$ (dark blue), $10^{-12}~{\rm GeV}^{-1}$ (light blue) and $10^{-10}~{\rm GeV}^{-1}$ (orange). Green curves show the case when there is no occurence of superradiance. (Top-right) Extra-galactic gamma-ray signal from Hawking radiation of the same benchmark points.
     (Bottom-left) Galactic line signals from the decay of the ALP cloud for the same benchmark points as in the top row. The black dashed curve shows the future sensitivity of SXI in the THESEUS mission. (Bottom-right) Galactic line signals in the eV region with sensitivity at JWST for an ALP cloud.}
    \label{fig:SignalSpectrum}
\end{figure*}

\noindent We show a benchmark example of multi-messenger signal spectra in the galactic and extra-galactic gamma-ray searches for PBH Hawking radiation, and the superradiated ALP decay line signals in the X-ray and infrared energies.

In Fig.~\ref{fig:SignalSpectrum} (top-left), we display the galactic center gamma-ray spectrum from Hawking radiation for a PBH of $M_{\rm{PBH}} = 6\times10^{16}~{\rm g}$, corresponding to the benchmark point in Fig.~\ref{fig:Ninfga}, with a dark matter fraction of $f_{\rm PBH} = 8\times10^{-5}$ (orange line) and $f_{\rm PBH} = 2\times10^{-3}$ (light and dark blue lines). The green line corresponds to no superradiance. The initial spin of the PBH for the lines is $a_{\ast,i}=0.99$. The most important effect of superradiance on Hawking radiation is the reduction in the radiation power after PBHs lose their spin to the production of ALPs. The wavy nature of the lines are due to contributions from various $l$ modes in Hawking radiation.  
The orange line corresponds to $g_{a\gamma\gamma}=10^{-10}$ GeV$^{-1}$ and the dark (light) blue lines are for $g_{a\gamma\gamma}=10^{-15}$ ($10^{-12}$)~GeV$^{-1}$. Larger $g_{a\gamma\gamma}$ values lead to quenched superradiant growth rates allowing a higher Hawking radiation flux due to the lack of spin loss. We also show the current COMPTEL and projected AMEGO-X continuum gamma-ray constraints in the parameter space (there are competitive bounds in the $M_{\textnormal{PBH}}<10^{17}~\textnormal{g}$ region from $e^{\pm}$ final states~\cite{Boudaud:2018hqb,Su:2024hrp,Huang:2024xap}, but here we are focused on final state photons). We find that values of $g_{a\gamma\gamma}\lesssim 6.5\times 10^{-11}~{\rm GeV}^{-1}$, lower than the current globular cluster constraints, can be explored at AMEGO-X for ALP masses of $m_a\sim$ keV.
In Fig.~\ref{fig:SignalSpectrum} (top-right), we show the AMEGO-X forecast and current COMPTEL sensitivities to extra-galactic gamma-ray signals in the same PBH and ALP parameters. The extra-galactic signal at AMEGO-X could probe values of $g_{a\gamma\gamma}$ below the current globular cluster constraints.

In Fig.~\ref{fig:SignalSpectrum} (bottom-left), we show the flux of decay line signals from the superradiance cloud having ALPs with $m_a$= 2 keV and velocity dispersion $\sigma_v=160~{\rm km}/{\rm s}$. We find that the SXI sensitivity of $g_{a\gamma\gamma}$ can detect ALP coupling significantly below the stellar cooling bound with a minimal fraction of PBH as DM. The larger values of $g_{a\gamma\gamma}$ produce stronger signals due to prompt ALP decay. $M_{\textnormal{PBH}}$ and $f_{\textnormal{PBH}}$ values for the various colored lines are the same as the top figures. The bottom-right figures show the ongoing JWST sensitivity for $m_a= 1$ eV, $M_{\textnormal{PBH}}=2\times 10^{21}$~g, $f_{\textnormal{PBH}}=1$. The red and purple lines are for $g_{a\gamma\gamma}=10^{-10}$ and $10^{-11}$ GeV$^{-1}$, respectively. There is no Hawking radiation constraint corresponding to this line signal since the PBH mass of $\sim 10^{21}$~g is outside the mass range of currently evaporating PBHs. However, one can constrain this region using microlensing data.

\bibliography{main}

\end{document}